\documentclass[letterpaper]{sig-alt-release2}

\usepackage{amsmath}
\usepackage{amssymb}
\usepackage{aas_macros}
\usepackage{graphicx}
\renewcommand\vec[1]{\boldsymbol{#1}}
\newcommand{\grad}{\vec{\nabla}}

\begin{document}
\conferenceinfo{SCG'12,} {June 17--20, 2012, Chapel Hill, North Carolina, USA.}
\CopyrightYear{2012}
\crdata{978-1-4503-1299-8/12/06}
\clubpenalty=10000
\widowpenalty = 10000

%\begin{description}
%\item[Title] 
\title{The Sticky Geometry of the Cosmic Web}

\numberofauthors{5}
\author{
\alignauthor Johan Hidding\\
%	\affaddr{Kapteyn Astronomical Institute,}\\
	\affaddr{Kapteyn Institute,}\\
	\affaddr{Groningen University}\\ 
%	\affaddr{P.O. Box 800, 9700 AV, Groningen, the Netherlands}\\
	\affaddr{Groningen, the Netherlands}\\
	\email{hidding@astro.rug.nl}
\alignauthor Rien van de Weygaert\\
%	\affaddr{Kapteyn Astronomical Institute,}\\
	\affaddr{Kapteyn Institute,}\\
	\affaddr{Groningen University}\\
%	\affaddr{P.O. Box 800, 9700 AV, Groningen, the Netherlands}\\
	\affaddr{Groningen, the Netherlands}\\
	\email{weygaert@astro.rug.nl}
\alignauthor Gert Vegter\\
%	\affaddr{Johan Bernoulli Institute for Mathematics and Computer Science}\\
	\affaddr{Johan Bernoulli Institute}\\
	\affaddr{Groningen University}\\
%	\affaddr{P.O. Box 407, 9700 AK, Groningen, the Netherlands}\\
	\affaddr{Groningen, the Netherlands}\\
	\email{g.vegter@rug.nl}
\and
\alignauthor Bernard J.T. Jones\\
%	\affaddr{Kapteyn Astronomical Institute,}\\
	\affaddr{Kapteyn Institute,}\\
	\affaddr{Groningen University}\\
%	\affaddr{P.O. Box 800, 9700 AV, Groningen, the Netherlands}\\
	\affaddr{Groningen, the Netherlands}\\
	\email{bernard@astrag.demon.co.uk}
\alignauthor Monique Teillaud\\
%	\affaddr{Geometrica, INRIA Sophia Antipolis}\\
	\affaddr{INRIA}\\
	\affaddr{Sophia Antipolis, France}\\
	\email{monique.teillaud@inria.fr}
}
%\item[Authors] 
%Johan Hidding (Kapteyn Astronomical Institute, Groningen University),
%\item[and] 
%\begin{itemize}
%\item Rien van de Weygaert (Kapteyn Astronomical Institute)
%\item Gert Vegter (Johan Bernoulli Institute for Mathematics and
%Computer Science)
%\item Bernard J.T. Jones (Kapteyn Astronomical Institute)
%\item Monique Teillaud (Geometrica, INRIA Sofia-Antipolis)
%\end{itemize}
%\item[Email]
%johannes.hidding@gmail.com
%
%\item[Summary] %\end{description}
\date{June 2012}
\maketitle
\begin{abstract}
In this video we highlight the application of Computational Geometry to our 
understanding of the formation and dynamics of the Cosmic Web. The
emergence of this intricate and pervasive weblike structure of the Universe on 
Megaparsec\footnote{The main measure of length in astronomy is the parsec.
Technically a parsec is the distance at which we would see the distance
Earth-Sun at an angle of 1 arcsec on the sky. One Megaparsec is 3.26 million
light-years; this is approximately the distance between our Milkyway and the
nearest other galaxy: Andromeda.} scales can be approximated by a well-known
equation from fluid mechanics, the  Burgers' equation. The solution to this 
equation can be obtained from a geometrical formalism. We have extended and 
improved this method by invoking weighted Delaunay and Voronoi tessellations. 
The duality between these tessellations finds a remarkable and profound 
reflection in the description of physical systems in Eulerian and Lagrangian 
terms. 

The resulting Adhesion formalism provides deep insight into the 
dynamics and topology of the Cosmic Web. It uncovers a direct connection 
between the conditions in the very early Universe and the complex spatial 
patterns that emerged out of these under the influence of gravity. 
\end{abstract}

% A category with the (minimum) three required fields
%A category including the fourth, optional field follows...
%\category{D.2.8}{Software Engineering}{Metrics}[complexity measures,
%performance measures]

%\terms{Theory, Algorithms}


\section{The Cosmic Web}
The Cosmic Web is the fundamental spatial organization of matter on scales of a
few up to a hundred Megaparsec.  Galaxies, intergalactic gas and dark matter
exist in a wispy weblike arrangement of dense compact clusters, elongated
filaments, and sheetlike walls, amidst large near-empty void regions. The
filaments are the transport channels along which matter and galaxies flow into
massive high-density cluster located at the nodes of the web. The weblike
network is shaped by the tidal force field accompanying the inhomogeneous
matter distribution. 
%Largely defined by (proto)cluster peaks, it effects anisotropic gravitational
%collapse. 

Structure in the Universe has risen out of tiny primordial (Gaussian) density
and velocity perturbations by means of gravitational instability
\cite{Peebles1980,Zeldovich1970}. The large-scale anisotropic force field
induces anisotropic gravitational collapse, resulting in the emergence of
elongated or flattened matter configurations. According to Zeldovich, the
collapse of a primordial cloud (dark) matter passes through successive stages,
first assuming a flattened sheetlike configuration as it collapses along its
shortest axis. This is followed by a rapid evolution towards an elongated
filament as the medium axis collapses and, if collapse continues along the
longest axis, may ultimately produce a dense and compact cluster or halo
\cite{Shandarin1989}.  The hierarchical setting of these processes, occurring
simultaneously over a wide range of scales complicates the picture
considerably.

\section{Zeldovich \& Adhesion models}
The simplest model that describes the emergence of structure and complex 
patterns in the Universe is the Zeldovich Approximation (ZA) \cite{Zeldovich1970, Shandarin1989}. 
In essence, it describes a ballistic flow, driven by a constant (gravitational) 
potential. The resulting Eulerian position $x(t)$ at some cosmic epoch $t$ is 
specified by the expression, 
\[x(t) = q + D(t)\ u_0(q),\]
where $q$ is the initial ``Lagrangian'' position of a particle, $D(t)$ the
time-dependent structure growth factor and $u_0 = -\grad_{\!\!q} \Phi_0$ its
velocity. The nature of this approximation may be appreciated by the
corresponding source-free equation of motion, 
\[\frac{\partial u}{\partial D} + (u \cdot \grad_{\!\!x}) u = 0.\]
The use of ZA is ubiquitous in cosmology. One major application is its key role
in setting up initial conditions in cosmological N-body simulations.  
Of importance here is its 
nonlinear extension in terms of Adhesion Model. 

\begin{figure}[h]
\centering
\epsfig{file=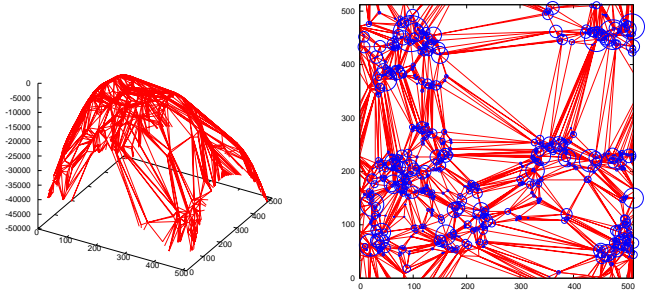, width=0.49\textwidth}
\caption{Convex hull of the modified potential and the resulting weighted
Delaunay triangulation.}
\end{figure}

\begin{figure}[h]
\centering
\epsfig{file=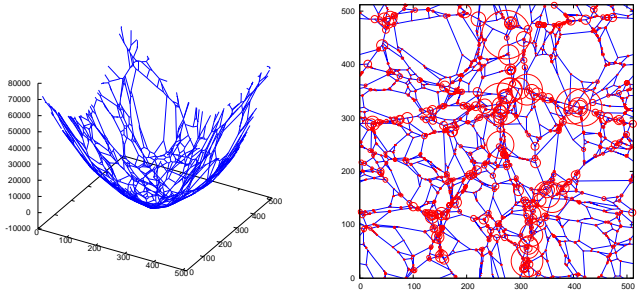, width=0.49\textwidth}
\caption{Convex conjugate of the modified potential and the resulting weighted
Voronoi diagram.}
\end{figure}

The ZA breaks down as soon as self gravity of the forming structures 
becomes important. To `simulate' the effects of self gravity, Gurbatov et 
al. \cite{Gurbatov1984, Gurbatov1989} included an artificial viscosity. 
This results in the Burgers' equation, 
\[\frac{\partial u}{\partial D} + (u \cdot \grad_{\!\!x}) u = \nu \grad_{\!\!x}^2 u,\]
a well known PDE from fluid mechanics. This equation has an exact analytical 
solution \cite{Hopf1950}, 
\[u(x, D) = \frac{\int dq\ \frac{x - q}{D}\ \exp[G/(2\nu)]}
{\int dq \exp[G/(2\nu)]},\]
where $G(x, q, D) = \Phi_0(q) - (x - q)^2 / (2D)$. In the limit of $\nu
\to 0$, the solution is
\[\Phi(x, D) = \max_q\left[\Phi_0(q) - \frac{(x - q)^2}{2D}\right].\]

This leads to a geometric interpretation of the Adhesion Model. The 
solution follows from the evaluation of the convex hull of the
velocity potential modified by a quadratic term \cite{Vergassola1994}. 
We found that the solution can also 
be found by computing the weighted Voronoi diagram \cite{Okabe2000} of a 
mesh weighted with the velocity potential. It is of special importance
that this computation is done on a periodic space. 
This is essential to treat boundaries correctly.

\section{Adhesion \& Tessellations}
The duality between the weighted Voronoi diagram and the weighted Delaunay
triangulation has a deep connection with the concepts of Eulerian and
Lagrangian coordinates in fluid mechanics. The Voronoi diagram represents
volume.
A vertex of the Delaunay triangulation represents a mass-element. Its dual
Voronoi cell is the volume this mass-element occupies. 
As time moves on, the particle is expanding in a struggle for dominance over
its neighbours, a struggle that is lost once the vertex becomes redundant 
%As time moves on,
%the particle is expanding in a struggle for dominance over its neighbours.  
%A struggle that is lost once the vertex becomes redundant 
(in the sense that
it no longer contributes to the Delaunay triangulation).  
Physically the particle is
then trapped in a larger structure. In general particles first stream into
walls (objects of co-dimension 1), moving down into objects of increasing
co-dimension. \cite{Hidding2010,Hidding2012}.
%This process can be better understood if we study it from the
%dual points of view of the Voronoi versus Delaunay diagrams.

%Redundant points in this triangulation correspond to particles
%that have collapsed under their own gravity. 
\section{This video}
We use the Computational Geometry Algorithms Library
\cite{cgal} for our key geometric computations: 
weighted Delaunay triangulations in both 2D and 3D (at this point, the periodic
triangulation package of CGAL only offers 3D periodic non-weighted Delaunay
triangulations).
%of weighted Delaunay triangulations in both 2D and 3D.  
%Also we used the unweighted Delaunay triangulation and alpha-shapes
%from the same library, in our analysis of observational data.

This video takes us from the initial conditions, through both the Zeldovich and
Adhesion models, to a display of how the topology of the Cosmic Web depends
directly on the starting point.

%\section*{References}
\bibliographystyle{abbrv}
\bibliography{film}{}

\end{document}